# THE EVOLUTION OF CYBERINSURANCE

**Ruperto P. Majuca**[†]     **William Yurcik**[‡]     **Jay P. Kesan**[φ]

[†] Department of Economics;
[‡] National Center for Supercomputing Applications (NCSA)
[φ] College of Law

University of Illinois at Urbana-Champaign
<majuca@uiuc.edu>  <byurcik@ncsa.uiuc.edu>  <kesan@law.uiuc.edu>

**Abstract.**    Cyberinsurance is a powerful tool to align market incentives toward improving Internet security. We trace the evolution of cyberinsurance from traditional insurance policies to early cyber-risk insurance policies to current comprehensive cyberinsurance products. We find that increasing Internet security risk in combination with the need for compliance with recent corporate legislation has contributed significantly to the demand for cyberinsurance. Cyberinsurance policies have become more comprehensive as insurers better understand the risk landscape and specific business needs. More specifically, cyberinsurers are addressing what used to be considered insurmountable problems (e.g., adverse selection/asymmetric information, moral hazard, etc.) that could lead to a failure of this market solution. Although some implementation issues remain, we suggest the future development of cyberinsurance will resolve these issues as evidenced by insurance solutions in other risk domains.

## 1. Introduction

As organizations become more dependent on their networked computer assets, the more vulnerable they become to harm from increasing frequent and damaging attacks enabled by connectivity. Protection from harm on any networked computer system will never be 100%. In the past decade, protection techniques from a variety of computer science fields such as cryptography and software engineering have continually made improvements and yet Internet attacks continue to increase (CERT/CC, 2005). While the mass media focuses on high profile Internet attacks in the form of security breaches and fast-spreading worms, undetected and unreported insider attacks from people within an organization with privileged access actually occur with greater frequency and impact (CSI/FBI, 2005). While some/most Internet security vendors are selling solutions in the form of hardware and software, Internet security protection is a continual process involving people that cannot be solved entirely with products (Schneier, 2000). Most relevantly, while most organizations have focused on preventing cyberattacks solely by technical means, this is only part of an overall solution. An overall solution must include accepting and managing the risk from cyberattacks since their occurrence is a reality.

A small group of interdisciplinary thinkers has proposed using cyberinsurance as part of the overall solution to Internet security. The earliest work describing a distributed systems application of insurance to the Internet dates back to 1994 (Lai et al., 1994). Dan Geer has been a prophet for the use of risk management, including insurance for the Internet. He was the first to state the relevance of risk management as commonly used in other fields especially the financial sector (Geer, 2003; Blakely et al, 2002; Geer 1998). The most eloquent spokesperson who brought cyberinsurance into academic discussion was Bruce Schneier who outlined what have now become consensus views about the role of cyberinsurance (Schneier, 2002; Schneier, 2001). Other commentators, including ourselves, have provided more details about cyberinsurance (Böhme, 2005; Kesan et al, 2005b; Ogut et al., 2005; Kesan et al, 2004; Yurcik and Doss, 2002; Varian, 2000). In 2002 we presented the role of insurance carriers in improving cybersecurity (Yurcik and Doss, 2002). Ideally cyberinsurance increases Internet safety because the insured increases self-protection as a rational response to the reduction of premium. Cyberinsurance also facilitates standards of liability. In 2004 we presented the economic theory and calculated welfare benefits for cyberinsurance (Kesan et al, 2004). In 2005 we analyzed several cyberinsurance policies and compared economic theory with the reality of what was happening in the marketplace (Kesan et al, 2005b).

In this paper, we look at the development of cyberinsurance over time since when it was first introduced in the late 1990s through 2005. Cyberinsurance development has certainly been affected by unforeseen events that we discuss in this paper. While it is hard to make generalizations over a market consisting of specific cyberinsurance contracts, we do identify trends that have occurred which may reveal insights for the future. The remainder of this paper is organized as follows: Section 2 presents the rational self-interest for cyberinsurance from complementary business perspectives. Section 3 traces the development of cyberinsurance from traditional insurance policies that were inadequate to early hacker insurance policies that developed into more comprehensive cyberinsurance policies. Section 4 discusses the problems cyberinsurers face and the mechanisms they are adopting to solve these problems. We end with a summary and conclusions in Section 5.

## *2. Business Perspectives on Cyberinsurance*

There are two business perspectives on cyberinsurance: (1) the insurer who seeks to capture profit from premiums exceeding losses over time by spreading the risk of uncertain loss events across many independent clients and (2) the individual or organization who seeks to maximize their utility/profit by managing the risk of uncertain loss events.

From the insurer perspective, cyberinsurance represents a growth opportunity since there is a growing need to protect core assets such as network infrastructure, data, and reputation. If an insurance firm can more accurately quantify the cyber-risks into attractive premiums, this opportunity may translate into a profit windfall. If premiums are priced too high then other insurers will reap the windfalls. However, if an insurance firm is inaccurate in quantifying the cyber-risks in premiums that are priced too low, then large losses may result.

Quantifying cyber-risks for the optimal premium price point is a difficult task since the assets to be protected are largely intangible, risk changes occur quickly (zero day threats appear suddenly), and evaluating the insurability of potential clients while re-evaluating risks with current clients can be resource intensive. However, balancing cost and risks is something the insurance industry has been doing for centuries. Brokers have emerged to arbitrage the business opportunity by matching customers with insurance underwriters.

Beyond determining the premium price point for different cyberinsurance policies, insurers are also faced with the critical challenge of spreading risk across many independent clients. Unfortunately, for cyberinsurance many of the recent Internet worm and virus attacks have had worldwide effects such that it is difficult to find clients whose risks are not dependent (Böhme, 2005; Ogut et al., 2005). An insurer may seek to spread risk over different hardware and software platforms, large and small organizations, etc. but it is an open question if risk can be partitioned on the Internet due to its innate connectivity.

From the individual or organization's perspective, the uncertainty of cyber-risks represents real risk for damages. There are four options for managing these risks:

1) avoiding the risk
2) retaining the risk
3) mitigating the risk
4) transferring the risk for a fee.

The first option is to avoid being exposed to cyber-risks by not having any dependence on computers, networked machines, or any Internet website presence. For some people/organizations this is feasible, however, for most commercial organizations this is not economically possible. The second option is to retain the risk based on a conscious decision that it is more cost effective to absorb any loss internally or other risk management options are unaffordable. A person or organization may place this bet based on informed judgment or risk-seeking behavior. Unfortunately, retaining the risk is sometimes the only choice due to lack of financial resources. The third option is to mitigate risk using managerial and technical processes. This involves investment in people and devices to identify threats and prepare counter-measures with continually improving security processes. While this option has been the exclusive focus of computer security professionals for decades, note it is just one of the risk management options. The fourth option is to transfer risk to a third party for a fee, in which case this third party must be licensed as an insurance company for performing this function. Insurance allows an individual or organization to smooth payouts for uncertain events (variable costs) into predictable periodic costs.

Typically an individual or organization employs a combination of these risk management options simultaneously – retaining some of the risk, mitigating some of the risk, and insuring the rest of the risk (Schneier, 2001). For example, a firm may choose to have an Internet website protected by security processes but yet avoid the risks of certain Internet transactions. One increasingly common risk management approach is to retain all or most of the risk while transferring the risk mitigation function to a third party (outsourcing) due to the superior expertise and cost efficiencies of the third party. Another common risk management hybrid is transferring specific risks via a product warranty or service contract. While in this paper we focus on cyberinsurance, it should be viewed as just one of several complementary risk management options – albeit the option we posit is growing in importance.

A risk management approach is commonly accepted in many industries – protection is relative and risks must be managed depending on the preferences of the parties involved and tradeoffs due to the specific environment. This is different from the prevailing approach in the IT industry for many years where protection is viewed as absolute and risks are to be avoided – computer scientists developing technology to "solve" Internet security problems. This mindset is slowly changing with the realization that absolute security is impossible and too expensive to even approach while adequate security is "good enough" to enable normal functions – the rest of the risk that cannot be mitigated can then be transferred via an insurance product.

Combining the two perspectives of insurers and individuals/organizations together, the primary business logic of cyberinsurance is as follows:

- As Internet connectivity increases the vulnerability of organizations to damages, organizations seek to manage this risk using cyberinsurance as one option in concert with other risk management options.

- Cyberinsurers recognize the opportunity to profit from the cyberinsurance risk management option and offer policies while simultaneously developing standards for insurability. Insurers are driven to find the best metrics in order to define profitable price ranges for different coverages given supply and demand.

- Organizations that manage risk using cyberinsurance as one option have increasing economic incentives to reduce exposure in tangible ways, for example by following 'best practices' specification of the techniques and equipment to be used for security protection. The economic incentives take two primary forms: (1) lower insurance premiums (discounts) for better security protection and (2) financial oversight which increasingly requires individuals/organizations to demonstrate that they protect their networked resources (considered part of fiduciary duty for executives in most commercial organizations).

- The end result is a market-solution with aligned economic incentives between cyberinsurers and individuals/organizations. Cyberinsurers seek profit opportunities from accurately pricing cyberinsurance and individuals/organizations seek to hedge potential losses.

There is more complexity beyond this primary business logic so we briefly note these secondary effects for completeness. This logic is premised on the competitiveness of the market structure, other market structures (monopoly, oligopoly) may change these dynamics. Since Internet security is interdependent, individuals/organizations may under-invest in security protection in order to free-ride on the investment of others. Since information about the level of security protection is often known completely only by the individual or organization, insurers and others may have to make decisions under uncertainty. Externalities and information asymmetries are addressed later in this paper.

## 3. *Development of Cyberinsurance*

*3.1 Traditional Insurance Policies*

Traditionally, firms rely upon several insurance policies to cover their losses from business: (1) business personal insurance policies (to cover first-party losses); (2) business interruption policies; (3) commercial general liability (CGL) or umbrella liability insurance policies (to cover liability for damages to third parties); and (4) errors and omissions insurance (to cover the firm's officers) (Lee, 2001). These traditional insurance policies were designed to cover the traditional perils of fires, floods, and other forces of nature. Since they were written before the advent of the Internet, they do not expressly cover new Internet risks. This has resulted in costly litigation between insurers and their policyholders; insurers drafting more ironclad exclusions (Duffy, 2002); and insurers developing new insurance policies to prevent inclusion of cyber-losses (Beh, 2002).

As an example, because cyber-properties do not necessarily have a physical form, attacks on them may not result in any physical damage. Accordingly, many disputes have arisen between insurers and firms as to what constitutes "tangible" property and "physical" damage as contemplated by the wording of the traditional policies. Thus, in Retails Systems, Inc. v. CNA Insurance Companies, 469 N.W.2d 735 [Minn. App. 1991], the court ruled that computer taps and data are tangible property under the CGL since the data had permanent value and was incorporated with the corporeal nature of the tape. Also, in American Guarantee & Liability Insurance Co. v. Ingram Micro, Inc., Civ. 99-185 TUC ACM, 2000 WL 726789 (D. Ariz. April. 18, 2000), the Arizona court ruled that the loss of programming in a computer's RAM constituted physical loss or damage. So too, in Centennial Insurance Co. v. Applied Health Care Systems, Inc. (710 F.2d 1288) [7th Cir. 1983], the court ruled in favor of the insured in a dispute concerning defective data processing and system failure which resulted in data loss. However, in Lucker Mfg. v. Home Insurance (23 F.3d 808 [3d Cir. 1994]), the Third Circuit ruled that the insured's liability for the loss of design use was not loss of tangible property use. So also, in Peoples Telephone Co., Inc. v. Hartford Fire Insurance Co., 36 F. Supp. 2d. 1335 [S.D. Fla. 1997] the Florida District Court ruled that electronic serial numbers and mobile telephone identification numbers are not tangible property.

Additionally, another problem is that while most CGL policies do not have worldwide coverage, many cyber-torts are international (Crane, 2001). If a firm's insurance policy stipulates risk coverage, it is uncertain if this encompasses international torts (Crane, 2001). Consequently, the inability of traditional insurance to deal with new cyber-threats creates the need to develop insurance products specifically designed to cover the new Internet risks.

*3.2 The Advent of Early Hacker Insurance Policies*

Although specialized coverage against computer crime first appeared in the late 1970s, these policies were an extension of the traditional crime insurance to electronic banking, and designed mainly to cover against an outsider gaining physical access to computer systems. It was not until the late 1990s that hacker insurance policies designed the Internet first appeared. The earliest known hacker insurance policies were first introduced in 1998 by technology companies partnering with insurance companies to offer clients both the technology services and first party insurance to either back up the technology company's technology or to provide a comprehensive total risk management solution to client firms (Nelson,1998; Davis, 1998; Duvall, 1998).

Being a new and unexplored area, these companies started out with small coverage. Thus, the International Computer Security Association (ICSA), the earliest group known to have offered hacker-related insurance as sort of warranty that its service is reliable, started out with only $250,000 maximum coverage per year.

Furthermore, almost all of these early hacker insurance policies covered only the insured firm's own (first party) loss. Table 1 illustrates how early hacker insurance started from simple and small amount coverage from losses against hacker attacks, to more differentiated products.

**Table 1.** *Early Hacker Insurance Products*

| Year | Company | Description | Coverage |
|---|---|---|---|
| 1998 | ICSA TruSecure (Poletti, 1998; Nelson, 1998) | product warranty | $1^{st}$ party coverage: max $20K per incident; max $250K per year |
| 1998 | Cigna Corp/ Cisco Systems/ NetSolve (Moukheiber, 1998; Clark, 1998, Davis, 1998) | partnership of insurance/benefits company with technology firms; client must buy security assessment and monitoring services | $1^{st}$ party (hacker damage & business interruption); $10M |
| 1998 | J.S. Wurzler Underwriting (Bryce, 2001) | insurance broker | $1^{st}$ party |
| 1998 | IBM/Sedgwick (Duvall, 1998; Greenemeier, 1998) | partnership between technology company and insurance firm | $5-15M |
| 2000 | Counterpane/ Lloyd's of London (Harrison, 2000) | partnership of security company with Lloyd's insurance | $1^{st}$ party; $1-10M |
| 2001 | Marsh McLennan/AT&T (Salkever, 2002) | clients who purchase from AT&T Internet data center receive a discount from insurance company | $1^{st}$ party |
| 2000 | AIG (Greenberg, 2000) | start of more comprehensive and sophisticated forms of insurance | $1^{st}$ & $3^{rd}$ party (infringement, libel, slander, privacy, invasion, errors & omissions); $25M |

*3.3 Causal Events: Increasing Risks and Legislation Compliance*

Perception of risk changed dramatically on September 11$^{th}$, 2001. There had been many Internet security events prior to 9/11 but afterward risks have been considered differently. Three of the most serious Internet worm attacks took place during a three month period around 9/11 – Code Red in July 2001, Nimda in September 2001, and Klez in October 2001. The Slammer Internet worm appeared in January 2003. Prior to 9/11 in February 2000, a series of coordinated denial-of-service (DoS) attacks were launched against major US corporations. Not only did the attacks prevent 5 of the 10 most popular Internet websites from serving its customers but the attacks also slowed down the entire Internet - Keynote Systems measured a 60% degradation in the performance of the 40 other websites that had not been attacked (Nelson, 2000).

In addition to these attacks, hackers have also engaged in attacking authentication systems, computer intrusions, web defacements, phishing, and identity theft (Kesan and Majuca, 2005a). Surveys reveal that 90% of businesses and government agencies have detected security breaches, 75% of these businesses suffered a resulting financial loss, 34% of organizations admit of less-than-adequate ability to identify if their systems have been compromised, and 33% admit lack of ability to respond (Insurance Information Institute, 2003). In fact, crackers have intruded into not only businesses but even key government agencies such as the U.S. Senate, Federal Bureau of Investigation (FBI), the National Aeronautics and Space Administration (NASA), and the Department of Defense (DoD) (Vogel, 2002; Insurance Information Institute, 2003). The Love Bug virus (2000) affected 20 countries and 45 million users caused an estimated $8.75 billion in lost productivity and software damage (Insurance Information Institute, 2003). Clearly, Internet risks have increased during 2000-2003 resulting in a need for individuals and organizations to manage this increased risk.

Simultaneous with the increasing risk from Internet attacks has been regulation about the legal use and retention of electronic information. Much of this legislation was started with the need for updated standards given computerized records and then driven by large corporate fraud events (Enron and Worldcom). Table 2 summarizes the laws and regulations related to information usage and storage. Sarbanes-Oxley Act, HIPAA, Gramm-Leach-Bliley Act and others mandate that financial information, patient records, and other client-related information must be handled in a secure manner. Penalties include corporate, civil, and criminal sanctions, with individual accountability to the CEO level. To meet these responsibilities, risk management in the form of both mitigation and insurance is required. Firms affected by these laws are held to a higher standard. Commentators suggest that other firms not specifically covered by the regulations in Table 2 may have a general common law duty to protect the information under their control (Smendinghoff, 2005; Kiefer and Sabett, 2002; Raul, Volpe and Meyer, 2001; Kenneally, 2000).

**Table 2.** *Federal/State Legislation Pertaining to Electronic Records*

| Legislation | Description |
| --- | --- |
| Health Insurance Portability and Accountability Act (HIPAA) (1996) | Mandates confidentiality, integrity, and availability of patient and other medical records and requires healthcare providers to secure the stored records. |
| SEC 17a-3 and 17a-4 (revised 2002) | Requires financial entities to retain client correspondence and all electronic records preserving integrity for auditing. |
| Gramm-Leach-Bliley Act (1999) | Requires financial entities to disclose policies for protecting confidential customer information and ensure integrity and confidentiality while preventing unauthorized access. |
| Sarbanes-Oxley Act (2002) | Ensures accuracy and reliability of corporate disclosures by requiring validation of integrity and accuracy of financial records. |
| California State Law SB 1386 (2002) | Requires all state agencies and businesses that store client information to promptly disclose security breaches. |

The combination of increased risks and compliance requirements is shown in Figure 1. Insurance products specifically designed for the Internet matured from rudimentary early insurance policies prior to 9/11 to more sophisticated cyberinsurance products post-9/11. Although there are many intervening variables, we assert that increasing risks and compliance requirements are the primary causal factors affecting this change in the development of cyberinsurance.

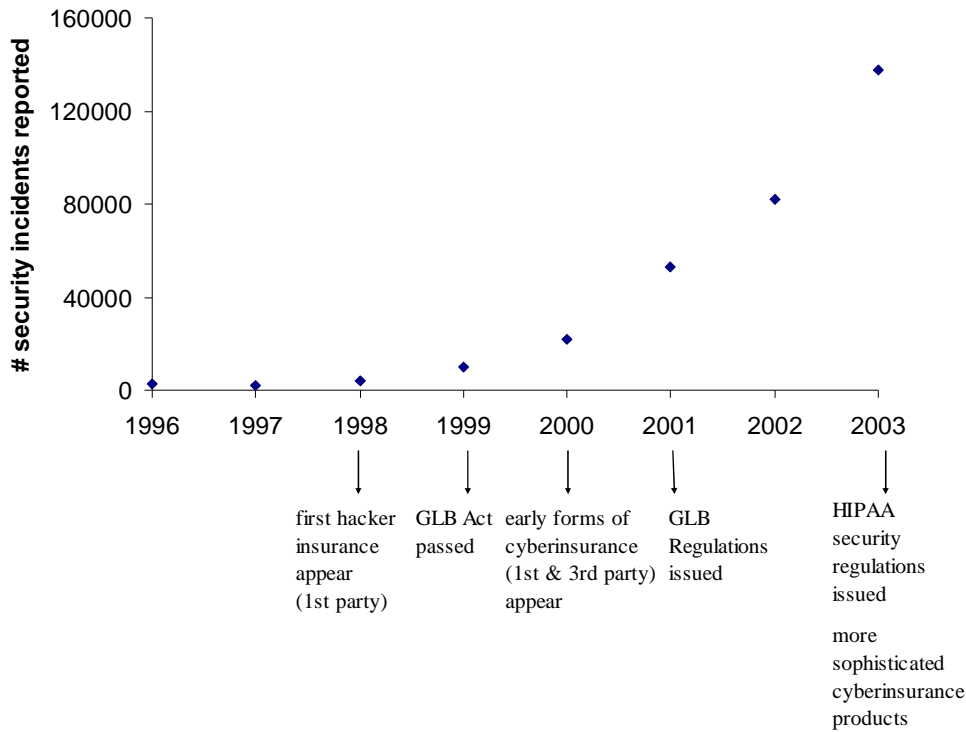

**Figure 1**. *Internet Incidents vs Relevant Laws and Cyberinsurance Products (data from CERT/CC, 2005).*

*3.4 More Sophisticated Cyberinsurance Policies*

Some examples of the new cyberinsurance products include American International Group (AIG) Inc.'s NetAdvantage, Lloyds of London's e-Comprehensive, and products InsureTrust.com, J.H. Marsh & McLennan, Sherwood, CNA, and Zurich North America (Wiles, 2003). Premiums can range from $5,000 to $60,000 per $1 million of coverage (or from 0.5% to 6%), depending on the type of business and the extent of insurance coverage.

As can be gleaned from Table 3 below, the recent cyberinsurance products have become more sophisticated compared to the early hacker insurance products. Unlike the first hacker insurance products which focused on first-party losses, recent cyberinsurance products cover both first party and third party insurance, and offer higher coverage. First party coverage typically cover destruction or loss of information assets, internet business interruption, cyberextortion, loss due to DOS attacks, reimbursement for public relation expenses, and even fraudulent electronic fund transfers. Third party coverage typically cover claims arising from Internet content, Internet security, technology errors and omissions and defense costs.

**Table 3.** *Summary Table of Recent Cyberinsurance Policies*

| COVERAGE | Net Advantage Security | e-Comprehensive | Webnet Protection |
|---|---|---|---|
| **First Party Coverages** | | | |
| Destruction, disruption or theft of info assets | Y | Y | Y |
| Internet Business Interruption | Y | Y | Y |
| Cyberextortion | Y | Y | Y |
| Fraudulent electronic transfers | N | Y | N |
| Denial of service attack | | Y | Y |
| Rehabilitation expenses | | Y | Y |
| **Third Party Coverages** | | | |
| Internet Content | Y | Y | Y |
| Internet Security | Y | Y | Y |
| Defense Costs | Y | Y | Y |

Net Advantage (AIG, 2005); E-Comprehensive (Loyd's of London, 2002); Webnet Protection (InsureTrust.com, 2003)

Another noticeable feature of recent cyberinsurance products is that they have narrow coverages designed to target different kinds of consumers. One reason for this practice is that insurers are able to exclude coverage of unforeseeable events by narrowly defining the insurance coverage (Baer, 2003). Another rationale is that by defining coverage more specifically, cyberinsurers are able to engage in product differentiation and thus offer their products to specific markets. For example, cyberinsurers have created products that are specifically meant to target firms concerned about damage to their own systems, products designed for firms who only want third party liability coverages, or products designed to cover media liability.

Table 4 provides an example of how cyberinsurers engage in product differentiation to capture different segments of the market. In Table 4 we see that AIG has offered different types of cyberinsurance products to capture different segments of the market with varying insurance needs. As an example, the enactment of HIPAA resulted in healthcare companies being specifically covered by liability legislation, and hence cyberinsurers have now designed cyberinsurance products specifically targeting this sector. Also some policies cover some specific risks (e.g. loss or claim associated with breach of patents or trade secrets, or bulletin boards), which other products exclude.

**Table 4.** *Different AIG Cyberinsurance Products Reveal Product Differentiation Strategy*

| COVERAGE \ Net Advantage Product | 1 | 2 | 3 | 4 | 5 | 6 | 7 |
|---|---|---|---|---|---|---|---|
| Network Security Liability | | | Y | Y | | Y | Y |
| Web Content Liability | Y | Y | Y | Y | | Y | Y |
| Internet Professional Liability | | Y | | Y | | | Y |
| Network Business Interruption | | | | | Y | Y | Y |
| Information Asset Coverage | | | | | Y | Y | Y |
| Identity Theft | | | Y | Y | Y | Y | Y |
| Extra Expense | | | | | Y | Y | Y |
| Cyber-extortion | | | Y | Y | Y | Y | Y |
| Cyber-terrorism | Y | Y | Y | Y | Y | Y | Y |
| Criminal Reward Fund | | | | | Y | Y | Y |
| Crisis Communication Fund | | | | | Y | Y | Y |
| Punitive, Exemplary and Multiple Damages | Y | Y | Y | Y | | Y | Y |
| Physical Theft of Data on Hardware/Firmware | | | Y | Y | | Y | Y |

AIG Product Name: 1 NetAdvantage; 2 NetAdvantage Professional; 3 NetAdvantage Commercial; 4 NetAdvantage Liability; 5 NetAdvantage Property; 6 NetAdvantage Security; 7 NetAdvantage Complete (AIG, 2005)

Firms who recently purchased new cyberinsurance products cite as among its advantages: (a) the ability to transfer risk to an insurer so they feel sheltered; (b) the capability to take fast action against a threat; (c) continuous monitoring by experts; and (d) expediency, since traditional insurance does not provide adequate protection against e-risks. Current industry estimates reveal a growing demand for cyberinsurance products, as well. In fact, the Insurance Information Institute estimates that cyberinsurance could become a $2.5 billion market by 2005 (Mader, 2002; Gohring, 2002).

## *4. How Cyberinsurers Worked Out Issues in Developing Coverage*

In developing cyberinsurance from the traditional insurance products to the early hacker insurance policies to where it is now, cyberinsurers had several important implementation issues to address. In this section we examine these implementation issues and the mechanisms cyberinsurers are using to deal with them.

### *4.1 Adverse Selection*

In an ideal world, parties to a contract have perfect information relevant to the decision. However, in many circumstances, one party may possess less than full information on the nature of the product being contracted. In insurance settings, these problems arise when insurers are unaware of whether an applicant is high-risk or low-risk. Since the applicant knows whether he/she is high-risk or low-risk while the insurer does not, there is an information asymmetry between them that leads to what is known in the economics literature as the adverse selection problem. When these situations arise, theory suggests that insurers would offer two types of contract: a low premium, low coverage contract designed to cover the low risk firms, and a high premium, high coverage contract to target the high-risk ones. In equilibrium, the high risk firms choose a contract that has full insurance coverage, while the low risk ones chose a contract that has only partial coverage. That is, the low risk firms suffer, because while the high risk firms get full coverage, low risk firms do not (Rothschild and Stiglitz, 1976). And since some firms (i.e., the low-risk firms) are not able to fully insure, the first best solution is not achieved. Only the second best solution – i.e., the best solution under information constraint – is feasible. Adverse selection problems therefore result in dissipative social welfare lost. Drawing from the international trade literature on measuring social welfare (Grinols and Wong, 1991; Grinols, 1984; Irwin, 2002; Bernhofen and Brown, 2003; Feenstra, 2003), the welfare loss due to adverse selection can be calculated for in dollar amounts.

We use Figure 2 to show this calculation of welfare loss. The market value of income is the *y*-intercept of the "budget line" tangent to the indifference curve, can be used as a measure of welfare. By comparing the market value of income in the first-best case with full cyberinsurance to the situation where asymmetric information lowers social welfare, we can provide dollar estimates of society's welfare loss due to the asymmetric information problem. If there are two types of insured in the economy (high risk and low risk ones), and if the insurer cannot distinguish between these two types, the insurer will offer contract $F_H$ (full insurance contract) to high risk applicants but will not be able to offer $F_L$ (full insurance) to low risk
8

applicants. In that case, the high risk applicants will have incentive to mimic the low risk applicants and purchase $F_L$ also. That is, the equilibrium solution must be such that the high risk firms have no incentive to imitate the low risk firms, and the low risk firms do not have incentive to present themselves as high risk firms. We know that this is characterized by the insurer offering two types of contract: high premium, high coverage contract $F_H$, which the high risk firms will purchase, and a low premium, low coverage contract $P$, which the low risk firms will purchase (Rothschild and Stiglitz, 1976).

Therefore, under asymmetric information, the first best solution – full insurance contract $F_L$ to low-risk applicants and full insurance contract $F_H$ to high-risk applicants – is not attainable. Instead, only the second best solution – where insurers offer partial insurance coverage $P$ to low-risk applicants and full insurance coverage $F_H$ to high-risk ones – is achieved. Consequently, social welfare is reduced with the inability of some firms (i.e., the low-risk firms) to fully insure. This welfare lost due to the adverse selection problem can be computed as the amount $A - A^p$ in Figure 2.

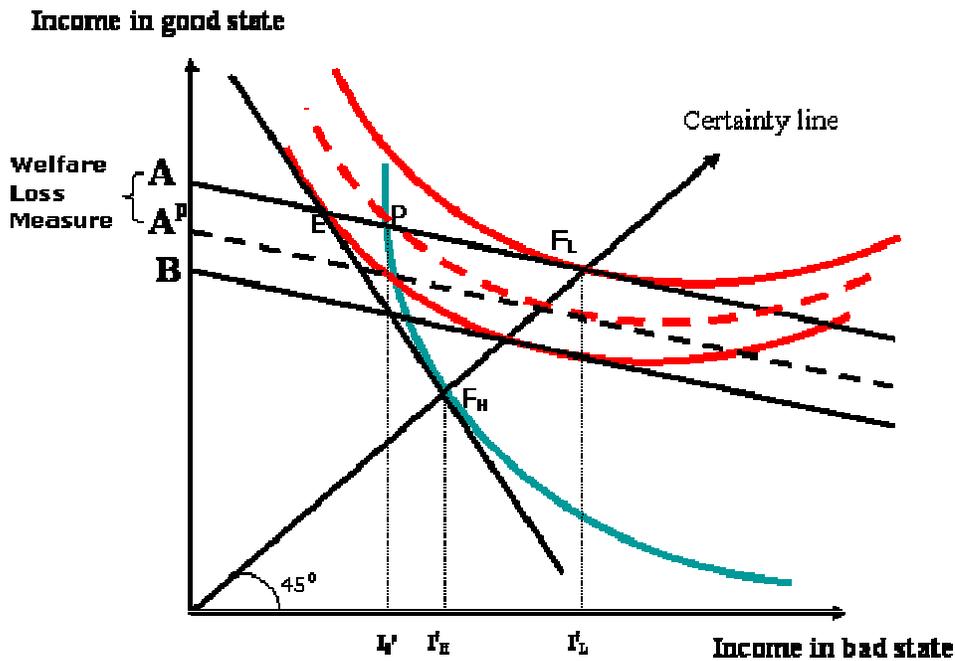

*Figure 2.* *Social Welfare Loss from Adverse Selection.*

To address the adverse selection problem, cyberinsurers require applicants to undergo thorough, detailed, and extensive risk assessments. As a condition to developing coverage, cyberinsurers evaluate the applicant's security through a myriad of offsite and on-site activities with a view of reviewing the applicant's vulnerabilities.

The risk assessment starts with the applicant filling in an application form with the detailed security questionnaire, some consisting of about 250 queries, to assess security risks and cyberprotections (technology budget, security infrastructure, virus-protection programs, testing and safety procedures, and outsourcing). General background questions include information on the applicant's Standard Industrial Classification (SIC) code; what Internet sites are proposed for insurance, including number of pages, customers/users, and page views; the annual sales and revenues, including revenue generated from Internet activities; IT budget and percentage of it earmarked for security; and what are specific Internet activities conducted (e.g., email and web browsing, production and internal processes integration, e-commerce, VPN, third party hosting services, consulting, etc.). More specific underwriting questions include information relating to:

<blockquote>

content: whether the applicant is monitoring its website's content; whether it has qualified intellectual property attorney and/or a written policy for removing controversial items;

what professional services are offered: whether the applicant's services include systems analysis, publishing, consulting, technology professional services, data processing, chatroom/bulletin boards, etc.; whether the applicant sells/licenses software or hardware; and whether there are hold and harmless clauses with subcontractors; and network security: whether there are company policies on IT security, privacy, and allowable email/Internet use; whether employees are informed of possible disciplinary actions for violation; whether third party security assessment and/or intrusion test were carried out; whether the high priority recommendations of the insurer were put into practice.

</blockquote>

The applicant needs to attach, among others, the firm's written policy on IT security, written policy for deleting offensive or infringing items, copy of appraisal of IT security controls and intrusion test outcomes, resumes of senior officers including the director of IT, and audited financial statements. Finally, the application form cites state laws reminding applicants that knowingly supplying false information is a crime in many states. This provides a direct incentive for applicants not to misrepresent their type of risk, at the risk of imprisonment (AIG, 2003; InsureTrust.com, 2001)

After examining the applicant's detailed application form, insurers then conduct a top-to-bottom physical and technical analysis of security, networks, and procedures. The baseline risk assessment starts with information requests on:

<blockquote>

physical security (including where equipment is located, single or multiple occupancy or multiple tenants, or whether the facility is a multi-story building, in a corporate campus or city, etc.)

network diagram (which shows the locations of operating systems, remote access devices, placement of routers, firewalls, web, database and email servers; which of systems reside in space leased from ISP; where each IP is located and what machines; and if hard drive or server space is leased); and description of network activities (e.g., list of IP addresses; list of managed devices like switches, hubs, routers, firewalls; platforms and OS including proxy servers, security scanners, anti-virus software, remote computer maintenance, main frame data protocols, firewall tunneling, wireless communications; etc.)

</blockquote>

Then follows physical reviews, including checks on applicant's personnel and hiring procedures, physical security review, review of incident response, disaster recovery, and security education programs, as well as technical assessment of the network's external vulnerability, using vulnerability scans, digital sweeps, network monitory for internal and external malicious users, and a review of firewalls, routers, network configuration. These results are analyzed and a report compiled listing recommendations for upgrades and fixes in order to ensure a more secure network (InsureTrust.com)

This is the mechanism cyberinsurers use to work around the adverse selection problem. The rigorous *ex ante* security assessment allowed insurers to distinguish between high and low risk applicants. By employing a clever mechanism of checking the applicants' security, insurers are able to avert a market failure that results from adverse selection and thus prevent the dissipation of social welfare arising from the asymmetric information problem. Furthermore, such mechanism works directly to benefit the low risk firms, since the security health checks enable them to distinguish themselves from the high-risk firms. With the ability of insurers to differentiate the risk types of applicants, high-risk applicants can no longer present themselves as low risk types and thus, the corresponding social welfare lost is averted.

*4.2 Moral Hazard*

The second major problem that insurers need to address in developing cyberinsurance coverage is the "moral hazard" problem. The problem is when firms are covered by insurance they may either intentionally cause the loss or take fewer measures to prevent the loss from occurring. For example, firms may slack in their security work when they are covered by insurance. Thus, they may either not invest in security infrastructure or they may not have incentive to maintain or upgrade their existing level of security.

The difference between the moral hazard problem and the adverse selection problem is in (1) costs and (2) their incentive structure. Addressing the adverse selection requires a sunk cost investment in decision-support infrastructure to determine risk classification of potential applicants that may not need to be revised very often. In contrast, the moral hazard problem requires investment in infrastructure to observe applicants that may need to be revised continuously. While the adverse selection problem deals with the incentive of the (high-risk) insured to hide information about its risk type to the insurer (that is why it is also called the *hidden information* problem in economics), the moral hazard problem deals with the incentive of the insured to slack in its action (thus it is also called the *hidden action* problem in economics). In insurance economics literature, a well-known device to work around the moral hazard problem is for insurers to observe the level of care that the insured takes to prevent the loss and tie the insurance premium to that amount of self-protection care. This way, the presence of insurance can in fact increase the level of self-protection that the insured takes rather than decrease it (Ehrlich and Becker, 1972; Shavell, 1979).

The security level can be perfectly observed either *ex ante* (before writing the insurance contract) or *ex post* (during the effectivity of the coverage), the presence of cyberinsurance increases the amount spent on self-protection by the insured firms as an economically rational response to the reduction of insurance premium, and thus results in higher levels of IT security in society. Thus, the detailed risk assessment conducted by insurers in developing cyberinsurance coverage works both to identify the risk type of the insured (thus address the adverse selection problem), and insofar as tying the risk classification to premium incentives the insured to adopt a higher level of security, it also addresses the moral hazard problem.

In examining current industry practice as well as several of the provisions of the cyberinsurance policies, we find that insurers are able to address the moral hazard problem by instituting several mechanisms in the cyberinsurance contract. By requiring applicants to undergo *ex ante* security assessment, cyberinsurers charge premiums according to risk classifications. Insurance coverage to firms with less cyberprotections, with a greater percent of its business online, or in a highly-regulated business subject to high penalties like financial firms, are considered to be higher risk (Mullin, 2002). Thus, a typical cyberinsurer would categorize an applicant firm into one of several risk classifications and tie the premiums to the level of the firm's security, giving discounts to firms that have superior security processes. Insuredotcom.com also places its applicant into 1 or 30 risk classifications. For instance, a new dot-com with no credit card transactions is categorized differently from Amazon.com (Banham, 2000). Insurers also utilize monitoring of the firm's security processes, third-party security technology partners, rewards for information leading to the apprehension of hackers, and expense reimbursement for post-intrusion crisis-management activities. For example, Safeonline may subcontract technology risk assessment to companies like IBM and others, Marsh uses Internet Security Systems (ISS) as its partners, and AIG's technology partners include IBM, RSA Security and Global Integrity Corporation.

*Ex post*, cyberinsurers also conduct surveys of insured's information infrastructure, either as part of regular annual surveys of the insurers premises, as part decision to continue and/or modify their coverage, or in processing of a loss or a claim. Several other provisions incorporated in the standard insurance policies are designed to address the moral hazard problem are shown in Table 5. First, insurers stipulate in the contract that they are not liable for losses or claims arising from the insured's failure to maintain a level of security equal to or superior to those in place at the inception date of the policy. That is, as an inducement to have good security, insurance policies stipulate that insured firms cannot claim payment for loss or claim associated with failure to take reasonable actions to maintain and improve their security. Thus, e-Comprehensive always include the following provision in its different coverages: "Provided always that the Insured Company maintain System Security levels that are equal to or superior to those in place as at the inception of this Policy" (Lloyd's of London, 2002). A similar provision can be found in the Webnet sample policy, thus: "You agree to protect and maintain your computer system and your e-business information assets and e-business communications to the level or standard at which they existed and were represented…" (InsureTrust.com, 2003). Second, insurers also explicitly state that no coverage will be given to firms who fail to back up their files. By unanimously excluding loss or claim based on failure to back-up from insurance coverage, cyberinsurers give insured firms incentives to regularly back-up their e-files. Third, once breach has occurred, insurers incentivize insured firms to mitigate the loss. For instance, under Lloyd's e-Comprehensive policy, expenditures incurred by the insured in employing the services of the underwriter's information risk group in order to mitigate the extent of the loss are expressly covered as a first party loss (Lloyd's of London, 2002). AIG's netAdvantage, on the other hand, include as part of its first-party coverage a criminal reward fund to be rewarded to individuals who give information resulting in conviction of the cybercriminal, while Webnet expressly covers investigative expenses incurred by the insured (InsureTrust.com, 2003). Also, Webnet requires the insured to "[n]otify the police if a law is broken" and to "[i]mmediately take all reasonable steps and measures necessary to limit or mitigate the loss, claim, or defense expenses" (InsureTrust.com, 2003). E-comprehensive also require in covering first party losses arising from malicious copying, recording, or sending of the insured's trade secret" that the insured should have "taken reasonable measures to prevent such copying, recording or sending of such Information" (Lloyd's of London, 2002).

*Table 5. Exclusions that address the Moral Hazard Problem in Recent Cyberinsurance Policies*

| EXCLUSIONS | Net Advantage Security | e-Comprehensive | Webnet Protection |
|---|---|---|---|
| Failure to back-up | Y | Y | Y |
| Failure to take reasonable steps to maintain and upgrade security | Y | Y | Y |
| Fraudulent, dishonest and criminal acts of insured | Y | Y | Y |
| Ordinary wear and tear of insured's info assets | Y | Y | Y |
| Claim arising out of liability to related parties | Y | Y | Y |
| OTHER RELEVANT PROVISIONS | | | |
| Retentions | Y | Y | Y |
| Liability Limits | Y | Y | Y |
| Criminal Reward Fund/Investigative Expenses Covered | Y | | Y |
| Services by Information Risk Group to mitigate the impact of 1st party loss, covered | | Y | |
| Representations Relied Upon | Y | Y | Y |
| Regular/Annual Surveys of Insured's Facilities | Y | Y | Y |

Net Advantage (AIG, 2005); E-Comprehensive (Loyd's of London, 2002); Webnet Protection (InsureTrust.com, 2003)

In the case where perfect observation of the insured firms' level of security is not possible, other incentive mechanisms designed to check the moral hazard problem are incorporated in standard cyberinsurance policies. Thus, for example, retentions and liability limits are designed to make the insured somewhat a co-insurer interested in preventing the occurrence of the lost (Shavell, 1979). Thus, the insured covers the first losses (retentions) as the insurance covers only amount over which the coverage will apply. Note also that the retentions generally apply to each loss. Other provisions designed to check on the moral hazard problem are the exclusion from coverage of losses and claims caused by fraudulent or dishonest acts committed by the insured, as well as claims arising out liability to related parties. Thus by observing the level of precaution by the insured, cyberinsurers are able to base a firm's insurance premium on the insured firm's investment in security processes, thereby creating market-based incentives for e-businesses to increase information security.

*4.3 Other Implementation Issues*

Internet security externalities arise because of interdependencies from interconnectivity. Computer systems have interdependent security such that an event on one system may affect all its peers even if they are under different administrative control. Thus, if malicious code penetrates a system through a compromised machine, it can use this machine as a platform for further attacks (Heal and Kunreuther, 2003). For example, if an individual or firm does not use anti-virus software, if infected it may propagate infections of other systems under different administrative control. Because of this possibility of aggregating cyber-risk exposures, a major concern in developing cyberinsurance coverage is the potential of single Internet security events causing damage to many policy holders simultaneously (Böhme, 2005; Ogut et al., 2005). Insurers have put in place several mechanisms designed to alleviate the problem of interrelated risks. As shown in Table 6, insurers may exclude events from coverage in order to protect themselves from large-scale losses associated with interrelated risks. For example, a common exclusion relates to losses due to failures of electric and telecommunication facilities. These exclusions are designed to shield insurers from exposure to a single event resulting in a large-scale failure.

*Table 6.* *Exclusions that address Externalities in Recent Cyberinsurance Policies*

| EXCLUSIONS | Net Advantage Security | e-Comprehensive | Webnet Protection |
|---|---|---|---|
| Inability to use or lack of performance of software programs | Y | Y | Y |
| Electric and telecommunication failures | Y | Y | Y |

Net Advantage (AIG, 2005); E-Comprehensive (Loyd's of London, 2002); Webnet Protection (InsureTrust.com, 2003)

Another problem with the developing cyberinsurance industry is the lack of underwriting standardization. Since Internet risks are complex, an assessment of a company's security can cost thousands of dollars. For example, AlphaTrust Corp.'s (insured by Insuretrust) security assessment costs about $20,000, while Marsh's security assessment cost $25,000 (Banham, 2000). Realizing that a detailed top-to-bottom physical analysis can be onerous for buyers, some insurers have simplified their underwriting procedures. For example, Insuredotcom.com developed an online questionnaire, while AIG adopted a three-level underwriting process -- online application, online assessment based on the questionnaire and a remote evaluation of the firm's security, and physical assessment (Banham, 2000).

Unlike traditional insurance where decades of information are available, there is little actuarial data to guide firms looking to minimize Internet risks (Gohring, 2002). Because insurers rely on measurements of predictability to forecast probable risk and set prices, the absence of actuarial data for Internet risks makes it more difficult to determine premiums (Martin, 2002; Walsh, 2001). Risk-metrics need to be developed to sharpen the estimation of risks. A possible step in the right direction is a closer partnership of insurance brokers with security service providers (Walsh, 2001). Another possibility is coordinating regulation and standardizing the policies for computer-related coverage with the help of the National Association of Insurance Commissioners (NAIC), a private, non-profit organization of insurance regulators (Lee, 2001). The Critical Infrastructure Protection Board (CIPB), established by President Bush in October 2001, has developed a partnership to pool government and insurance data in order to develop actuarial tables (Duffy, 2002).

## *5.0 Summary*

Cyberinsurance is an important tool for Internet security at two different levels: (1) it aligns economic incentives for insurers and individuals/organizations to manage risks for profit and (2) the aggregate self-interest of a cyberinsurance market results in increased social welfare gains (Kesan et al., 2005). We conduct a longitudinal study by tracing the evolution of cyberinsurance from traditional insurance policies to early cyber-risk insurance policies to current comprehensive cyberinsurance products. We conclude that the cyberinsurance industry has matured from primitive "hacker insurance polices" offering largely first-party policies with low coverages, to more sophisticated, product-differentiated policies offering first- and third-party insurance products with substantially higher coverages.

We also find that cyberinsurance companies are able to deal with implementation issues. For instance, insurers are addressing adverse selection and the moral hazard problem by rigorously classifying the risk level of the insured, and stipulating provisions on the care expected of the insured. We present a methodology for calculating the amount of social welfare loss that may be averted by addressing problems such as adverse selection.

We conclude that cyberinsurance products are making the Internet a safer environment because cyberinsurers are requiring businesses to minimize losses using economic incentives and individuals/organizations are increasingly seeing cyberinsurance in their own self-interest. Insurers can pool knowledge about risks, identify system-wide vulnerabilities, demand that the insured undergo prequalification audits, and adopt pro-active loss prevention strategies (Beh, 2002). This is similar to what has happened in other industries where insurance increased safety in fire prevention, aviation, boiler and elevators (Kehne, 1986). In addition to compliance with federal legislation for protecting networked infrastructure, federal subsidies are an additional option for encouraging firms to purchase cyberinsurance following NAIC's model regulations and guidelines for such areas as accident and health insurance, and the intervention of the government for such areas as floods and nuclear power plant accidents (Lee, 2001). Unfortunately, change in risk management actions often require events of great magnitude. We have already mentioned 9/11 in this paper as one such event. The New York City Triangle Shirtwaist Factory fire of 1911 is the one horrific event credited with instituting fire protection and fire insurance in the United States. Unfortunately it may take an Internet event of similar magnitude before cyberinsurance increases its penetration to higher levels but this is not a question of "if" but "when".